\documentclass[preprint,aps,showpacs,nofootinbib,preprintnumbers,amsmath,amssymb]{revtex4-1}
\usepackage{epsfig}
\usepackage{subfigure}
\usepackage{dcolumn}
\usepackage{bm}
\usepackage[usenames ,dvipsnames]{xcolor}
\usepackage{slashed}
\usepackage{graphicx,color}

\begin{document}
\title{Pentaquarks in semileptonic $\Lambda_b$ decays}

\author{Ping Zhou$^1$, Y.K. Hsiao$^{1,2}$ and C.Q. Geng$^{1,2}$}
\affiliation{
$^1$Chongqing University of Posts \& Telecommunications, Chongqing, 400065, China\\
$^2$Department of Physics, National Tsing Hua University, Hsinchu, Taiwan 300
}
\date{\today}

\begin{abstract}
In terms of ${\cal P}_{c1,c2}={\cal P}_c(4380,4450)^+$ as the hidden charm pentaquark states to consist of 
$c\bar c uud$, we study the semileptonic $\Lambda_b\to ({\cal P}_{c1,c2}\to)J/\psi p \ell^-\bar \nu_\ell$ decays.
In our discussion, while the main contribution to $\Lambda_b\to K^-J/\psi p$ is from 
the non-perturbative process via the doubly charmful $b\to c\bar cs$ transition,
we propose that the $\Lambda_b\to {\cal P}_{c1,c2}\to J/\psi p$ transitions are 
partly contribute to the $\Lambda_b\to K^-({\cal P}_{c1,c2}\to)J/\psi p$ decays,
in which the required $c\bar c$ pair is formed by the sea quarks, intrinsic charm, or both. 
We predict that ${\cal B}(\Lambda_b\to J/\psi p \ell^-\bar \nu_\ell)=
(2.04^{+4.82}_{-1.57},1.75^{+4.14}_{-1.35})\times 10^{-6}$ for $\ell=(e,\mu)$, 
which are about two orders of magnitude
smaller than the observed decay of $\Lambda_b\to p\mu\bar \nu_\mu$.
We also explore the angular correlations for the $J/\psi p$ and $\ell^-\bar \nu_\ell$ pairs. Our  results of the 
decay branching ratios and angular asymmetries in $\Lambda_b\to J/\psi p \ell^-\bar \nu_\ell$, accessible 
to the ongoing experiments at the LHCb, can be used to  improve the understanding of the hidden charm pentaquark states.

\end{abstract}

\maketitle
\section{introduction}
In the $J/\psi p$ invariant spectrum 
of the $\Lambda_b\to J/\psi p K^-$ decay, the two resonant states of 
${\cal P}_{c1,c2}={\cal P}_c(4380,4450)^+$ 
have been identified as
the hidden charm pentaquark states of 
$c\bar c uud$~\cite{Pc_LHCb,Pc_LHCb2,Aaij:2016ymb},
which are in accordance with the theoretical predictions of the pentaquark to consist of 
the $c\bar c$ pair~\cite{Wu:2010jy,Karliner:2015ina,Yuan:2012wz,Chen:2015loa}.
After the observation, the theoretical studies 
have been mainly concerning the pentaqurk structure~\cite{Chen:2016qju},
such as those examined in
the diquark-diquark-antiquark models~\cite{Wang:2015epa,
Maiani:2015vwa,Anisovich:2015cia,Ghosh:2015ksa},
taken as the baryon-meson bound states of 
$\Sigma_c \bar D^*$, $\Sigma_c^* \bar D^*$ and 
$\chi_{c1}p$~\cite{Chen:2015moa,Roca:2015dva,Meissner:2015mza,Burns:2015dwa,Lu:2016nnt}
or the genuine multiquark states~\cite{Mironov:2015ica}. On the other hand,
the data can be interpreted with the kinematical effects related to 
the so-called triangle singularity~\cite{Guo:2015umn,Liu:2015fea,Mikhasenko:2015vca},
which may be viewed to be opposite
the pentaquark scenario~\cite{Burns:2015dwa,Wang:2015pcn}.
The pentaquark productions,
such as the $b$-baryon decays~\cite{Li:2015gta,Hsiao:2015nna}, 
photo-productions~\cite{Wang:2015jsa,Kubarovsky:2015aaa,Karliner:2015voa} and
$\pi$-$p$ collision~\cite{Lu:2015fva}
have been also examined.

\begin{figure}[t!]
\centering
\includegraphics[width=2.2in]{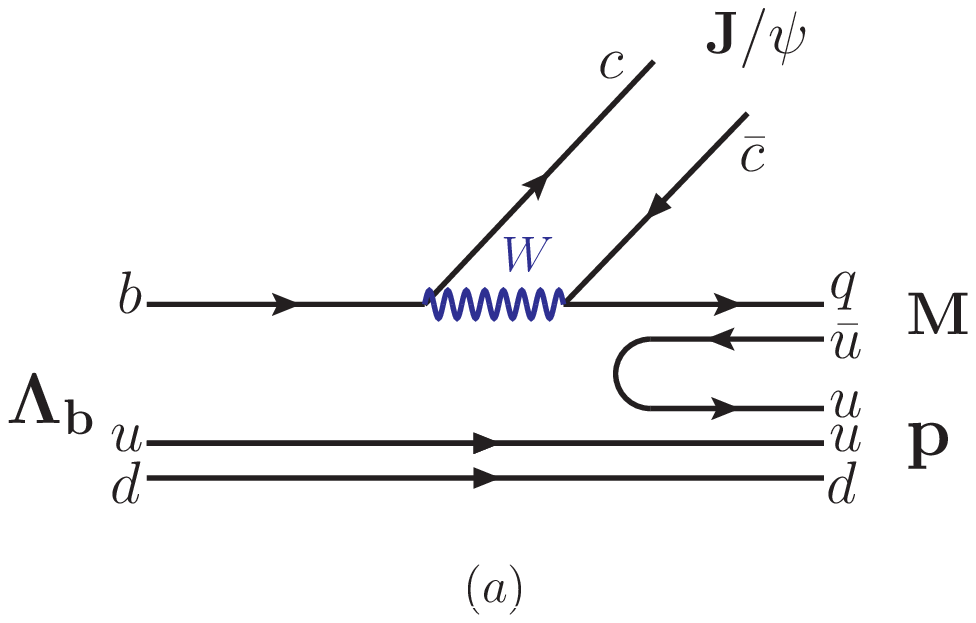}
\includegraphics[width=2.2in]{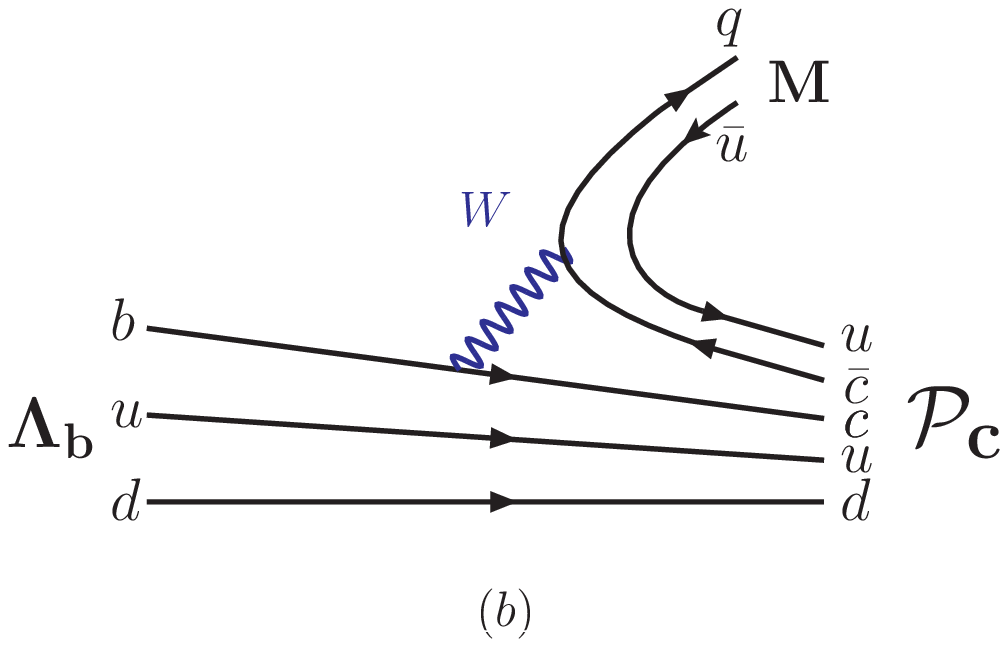}
\includegraphics[width=2in]{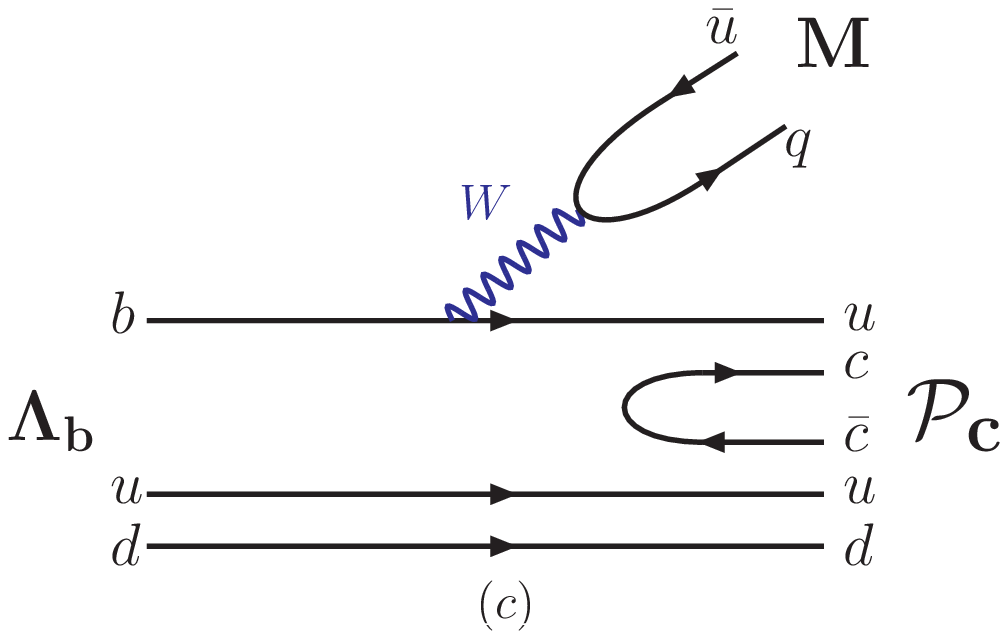}
\includegraphics[width=2in]{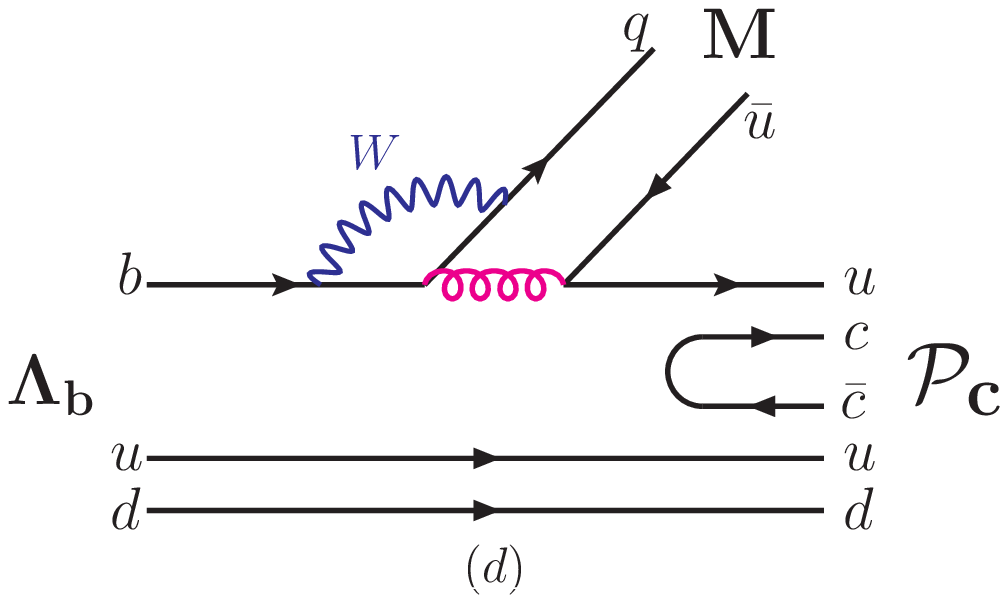}
\includegraphics[width=2in]{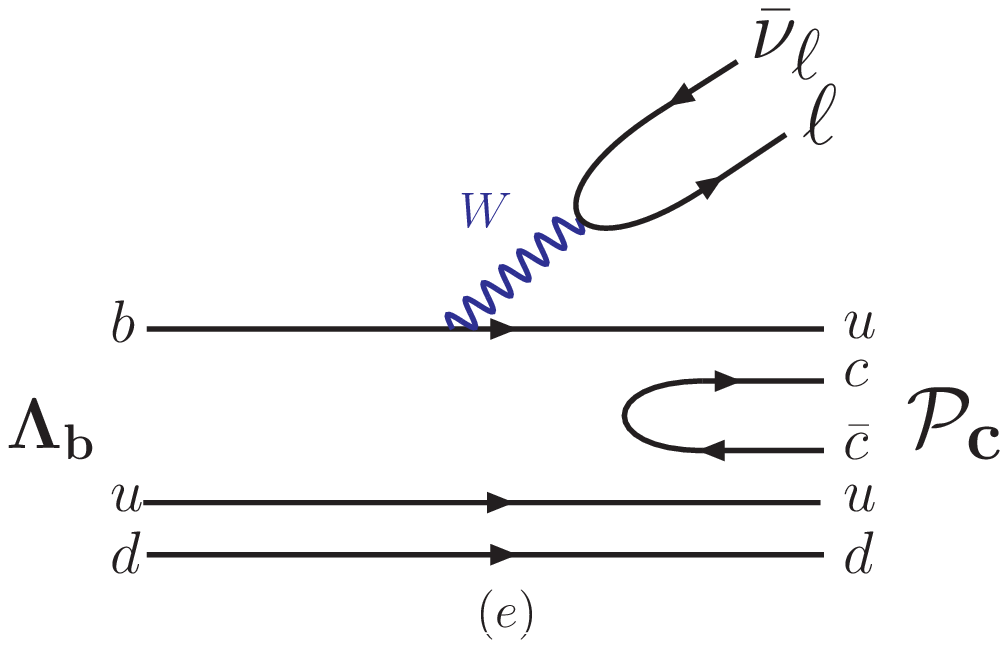}
\caption{
Three types of diagrams for 
the $\Lambda_b\to MJ/\psi p$ decays,
where
(a) is the first one without ${\cal P}_c$,
and (b) and (c,d) correspond to the second and third ones through
 the $\Lambda_b\to M{\cal P}_c(\to J/\psi p)$ processes
  via the doubly charmful $b\to c\bar c q$ 
  and $b\to u\bar u q$ transitions, respectively, while (e) is the diagram for 
the semileptonic $\Lambda_b\to {\cal P}_c(\to J/\psi p) \ell^-\bar \nu_\ell$ decays.}\label{dia}
\end{figure}

The $\Lambda_b\to M^-J/\psi p$ decays via the $b\to c\bar c q$ transitions
with $q=s(d)$ for $M=K(\pi)$ have two types of the diagrams as shown in
 Figs.~\ref{dia}a and \ref{dia}b. 
 The first one in Fig.~\ref{dia}a   
 proceeds with 
the direct $\Lambda_b\to p M$ and indirect $\Lambda_b\to N^*\to p M$ transitions,
accompanying the recoiled $J/\psi$ vector meson, with $N^*$ denoted as a higher wave baryon state.
This type of the diagrams 
makes the main contributions to the total decay branching ratios of $\Lambda_b\to (\pi^-,K^-)J/\psi p$. 
The second type is illustrated in Fig.~\ref{dia}b, which involves 
the non-perturbative processes with
the resonant $\Lambda_b\to K^-({\cal P}_{c1,c2}\to)J/\psi p$ decays 
as studied in Refs.~\cite{Liu:2015fea,Li:2015gta,Lebed:2015dca,Wang:2015pcn}.
On the other hand, 
$\Lambda_b\to M({\cal P}_{c1,c2}\to)J/\psi p$ 
can also be produced via the charmless $b\to u\bar u q$ transition~\cite{Hsiao:2015nna} as depicted in 
Figs.~\ref{dia}c and \ref{dia}d, in which
the required $c\bar c$ pair 
comes from 
the sea quarks (gluon splitting) or
the intrinsic charms within $\Lambda_b$ and 
$p$~\cite{IC1,IC2,Brodsky:1997fj,Chang:2001iy,Brodsky:2001yt,Mikhasenko:2012km}, 
along with 
the $\Lambda_b\to {\cal P}_{c1,c2}\to J/\psi p$ transitions and 
the recoiled $M$.
The current experimental  measurements on $\Lambda_b\to M^-J/\psi p$
by the LHCb Collaboration can be  summarized as follows:
\begin{eqnarray}\label{data1}
&&{\cal B}(\Lambda_b\to K^- J/\psi p)
=(3.17\pm 0.04\pm 0.07\pm 0.34^{+0.45}_{-0.28})\times 10^{-4}~\text{\cite{Pc_LHCb}}\,,
\nonumber\\
&&{\cal B}(\Lambda_b\to \pi^- J/\psi p)
=(2.61\pm 0.09\pm 0.13^{+0.47}_{-0.37})\times 10^{-5}~\text{\cite{Pc_LHCb}}\,,
\nonumber\\
&&{\cal B}(\Lambda_b\to K^-({\cal P}_{c1}\to)J/\psi p)=
(2.66\pm 0.22\pm 1.33^{+0.48}_{-0.38})\times 10^{-5}~\text{\cite{Aaij:2015fea}}\,,
\nonumber\\
&&{\cal B}(\Lambda_b\to K^-({\cal P}_{c2}\to)J/\psi p)=
(1.30\pm 0.16\pm 0.35^{+0.23}_{-0.18})\times 10^{-5}~\text{\cite{Aaij:2015fea}}\,,
\nonumber\\
&&\Delta {\cal A}_{CP}\equiv{\cal A}_{CP}(\Lambda_b\to J/\psi p \pi^-)-{\cal A}_{CP}(\Lambda_b\to J/\psi p K^-)
=(5.7\pm 2.4\pm 1.2)\%~\text{\cite{LHCb1}}\,,\nonumber\\
&&{\cal R}_{\pi K}\equiv \frac{{\cal B}(\Lambda_b\to J/\psi p \pi^-)}{{\cal B}(\Lambda_b\to J/\psi p K^-)}
=0.0824\pm 0.0025\pm 0.0042~\text{\cite{LHCb1}}\,,\nonumber
\end{eqnarray}
\begin{eqnarray}
&&{\cal R}_{\pi K}({\cal P}_{c1})\equiv 
\frac{{\cal B}(\Lambda_b\to {\cal P}_{c1} \pi^-)}{{\cal B}(\Lambda_b\to {\cal P}_{c1} K^-)}
=0.050\pm 0.016^{+0.026}_{-0.016}\pm 0.025~\text{\cite{Aaij:2016ymb}}\,,
\nonumber\\ 
&&{\cal R}_{\pi K}({\cal P}_{c2})\equiv 
\frac{{\cal B}(\Lambda_b\to {\cal P}_{c2} \pi^-)}{{\cal B}(\Lambda_b\to {\cal P}_{c2} K^-)}
=0.033^{+0.016+0.011}_{-0.014-0.010}\pm 0.009~\text{\cite{Aaij:2016ymb}}\,.
\end{eqnarray}

To understand the difference of the CP asymmetries in Eq.~(\ref{data1}), 
it is clear that the third type of the diagrams is needed as it is the 
only possible source to provide the weak CP violating phase from the CKM element of $V_{ub}$~\cite{Hsiao:2015nna}.
As a result, we have obtained~\cite{Hsiao:2015nna} that 
${\cal A}_{CP}(\Lambda_b\to \pi^-({\cal P}_{c1,c2}\to)J/\psi p)=(-7.4\pm 0.9)\%$, and 
${\cal A}_{CP}(\Lambda_b\to K^-({\cal P}_{c1,c2}\to)J/\psi p)=(+6.3\pm 0.2)\%$.
In addition, we have  shown that
${\cal B}(\Lambda_b\to \pi^-({\cal P}_{c1,c2}\to) J/\psi p)/
{\cal B}(\Lambda_b\to K^-({\cal P}_{c1,c2}\to) J/\psi p)=0.58\pm 0.05$ based on Figs.~\ref{dia}c and \ref{dia}d,
 whereas Fig.~\ref{dia}b via $b\to c\bar c q$ leads to that
${\cal B}(\Lambda_b\to \pi^-({\cal P}_{c1,c2}\to) J/\psi p)/
{\cal B}(\Lambda_b\to K^-({\cal P}_{c1,c2}\to) J/\psi p)
\simeq 0.07-0.08$~\cite{Li:2015gta}. 
While this ratio can be used to test 
which type of the diagrams is mainly responsible for the pentaquark productions,
the recent observations of 
${\cal R}_{\pi K}({\cal P}_{c1},{\cal P}_{c2})=(0.050\pm 0.039,0.033\pm 0.021)$
in Eq.~(\ref{data1})~\cite{Aaij:2016ymb} disfavor our assumption that 
the charmless $b\to u\bar u q$ transitions in Figs.~\ref{dia}c and \ref{dia}d
dominate the resonant $\Lambda_b\to M({\cal P}_{c1,c2}\to)J/\psi p$ decays. 
Nonetheless, apart from the nonfactorizable diagram in Fig.~\ref{dia}b, 
which is preferred by the data to give the main contribution,
the apparently large uncertainties of ${\cal R}_{\pi K}({\cal P}_{c1},{\cal P}_{c2})$ still
make the room for the other pentaquark productions as those in Figs.~\ref{dia}c and \ref{dia}d.

As shown in Fig.~\ref{dia}e, 
it is interesting to note that the $\Lambda_b\to {\cal P}_{c1,c2}$ transitions
proposed to explain $\Lambda_b\to K^-{\cal P}_{c1,c2}$ would simultaneously induce
the semileptonic $\Lambda_b\to {\cal P}_{c1,c2} \ell^-\bar \nu_\ell$ decays,
similar to $\Lambda_b\to p\ell^-\bar \nu_\ell$ 
associated with the $\Lambda_b\to p$ transition, with the only difference being that 
${\cal P}_{c1,c2}$ contain the additional 
$c\bar c$ pair compared to $p$. It is feasible that 
the semileptonic $\Lambda_b\to ({\cal P}_{c1,c2}\to)J/\psi p\ell^-\bar \nu_\ell$ decays
can be used to explore the resonant pentaquark states,
while the $B_c^-\to X(3872)\ell\bar \nu_\ell$ decays,  
proposed recently in Ref.~\cite{Wang:2015rcz,Hsiao:2016pml},
tackle the tetraquark state (the four-quark bound state).
Note that, as in Refs.~\cite{LHCb_Vub,Chen:2002zk,Gabbiani:2002ti},
the semileptonic $b$-hadron decays have been successfully
used to extract $V_{ub}$, test new physics, determine the form factors,
and seek the intrinsic charm. In this report, 
we will study the semileptonic $\Lambda_b\to ({\cal P}_{c1,c2}\to) J/\psi p \ell^-\bar \nu_\ell$ decays
to be connected with the pentaquark productions in the hadronic $\Lambda_b$ decays. 
Unlike the previous extraction of the information of 
the resonant $\Lambda_b\to({\cal P}_{c1,c2}\to)J/\psi p$ transitions in Ref.~\cite{Hsiao:2015nna}, 
which assumes that Figs.~\ref{dia}c and \ref{dia}d
dominate $\Lambda_b\to M({\cal P}_{c1,c2}\to)J/\psi p$, we will include 
the pentaquark contributions from Figs.~\ref{dia}b-\ref{dia}d 
as well as the new experimental inputs of ${\cal R}_{\pi K}({\cal P}_{c1},{\cal P}_{c2})$ 
to re-extract the information.
%
With the new extraction, we will then study
the $\Lambda_b\to ({\cal P}_{c1,c2}\to) J/\psi p \ell^-\bar \nu_\ell$ decay,
which can be clearly tested if there exist the third types of the contributions via $b\to u\bar u q$,
apart from the second ones via $b\to c\bar c q$, measured
as the main contributions of the resonant $\Lambda_b\to M({\cal P}_{c1,c2}\to)J/\psi p$ decays.

\section{Formalism}
In terms of the effective Hamiltonian of $b\to u\ell\bar \nu_\ell$ at the quark level from Fig.~\ref{dia}c,
given by
\begin{eqnarray}
{\cal H}(b\to u \ell^-\bar \nu_\ell)&=&\frac{G_F V_{ub}}{\sqrt 2}\;\bar u\gamma_\mu (1-\gamma_5)b\; 
\bar \ell\gamma^\mu (1-\gamma_5)\nu_\ell\,,
\end{eqnarray}
the amplitude for the $\Lambda_b\to ({\cal P}_c\to)J/\psi p \ell^-\bar \nu_\ell$ decay 
with $\ell=(e$ or $\mu$)
can be derived as
\begin{eqnarray}\label{amp}
&&{\cal A}(\Lambda_b\to ({\cal P}_c\to)J/\psi p\ell^- \bar \nu_\ell)=\frac{G_F V_{ub}}{\sqrt 2}
\langle J/\psi p|\bar u\gamma_\mu (1-\gamma_5)b|\Lambda_b\rangle \;
\bar \ell\gamma^\mu (1-\gamma_5) \nu_\ell\;.
\end{eqnarray}
In accordance with the pentaquark observations
in the resonant $\Lambda_b\to K^-({\cal P}_{c1,c2}\to) J/\psi p$ decays,
the matrix elements in Eq.~(\ref{amp})  for the resonant $\Lambda_b\to {\cal P}_c\to J/\psi p$ transition 
can be written as~\cite{Hsiao:2015nna}
\begin{eqnarray}\label{amp2}
\langle J/\psi p|\bar u\gamma_\mu(1-\gamma_5)b|\Lambda_b\rangle&=&
\langle J/\psi p|{\cal P}_c \rangle{\cal R}_{{\cal P}_c}
\langle {\cal P}_c|\bar u\gamma_\mu(1-\gamma_5)b|\Lambda_b\rangle\nonumber\\
&=&{\cal R}_{{\cal P}_c}(\varepsilon \cdot q) 
\bar u_p(\bar F_{{\cal P}_c}\gamma_\mu-\bar G_{{\cal P}_c}\gamma_\mu\gamma_5)u_{\Lambda_b}\,,
\end{eqnarray}
where 
$\varepsilon^{\mu}$ is the $J/\psi$ four-polarization,
$q^\mu$ is the momentum transfer,
$F_{{\cal P}_c}$ ($G_{{\cal P}_c}$) represents the vector (axial-vector) form factors,
and  ${\cal R}_{{\cal P}_c}$ corresponds to 
the Breit-Wigner factor,
given by
\begin{eqnarray}
{\cal R}_{{\cal P}_c}=\frac{i}{(t-m_{{\cal P}_c}^2)+im_{{\cal P}_c}\Gamma_{{\cal P}_c}}\,,
\end{eqnarray}
with $t=(p_{J/\psi}+p_{p})^2$.
As an intermediate state, 
the ${\cal P}_c$ spin, no matter 3/2 or 5/2, has been summed over,
included in $F_{{\cal P}_c}$ and $G_{{\cal P}_c}$.

\begin{figure}[t!]
\centering
\includegraphics[width=3.2in]{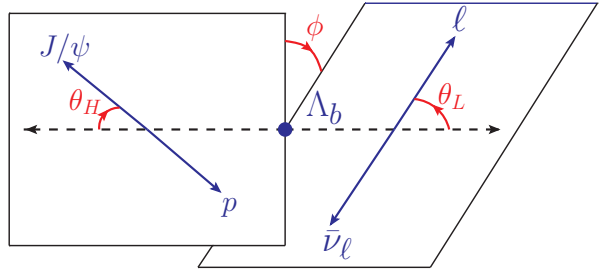}
\caption{The three angles $\theta_{H,L}$ and $\phi$ as the kinematic variables
for the semileptonic four-body $\Lambda_b\to J/\psi p \ell^-\bar \nu_\ell$ decay.}\label{dia2}
\end{figure}

To describe the four-body $\Lambda_b(p_{\Lambda_b})\to 
J/\psi(p_{J/\psi})p(p_p)\ell^-(p_\ell)\bar\nu (p_{\bar\nu})$ decay,
we need five kinematic variables: $s=(p_{\ell}+p_{\bar\nu})^2$,  $t=(p_{J/\psi}+p_{p})^2$,  
and the three angles $(\theta_H,\theta_L,\phi)$
shown in Fig.~\ref{dia2}~\cite{Kl4,Wise}. 
As seen in Fig.~\ref{dia2}, 
the angle $\theta_{H(L)}$ is between $\vec{p}_{J/\psi}$  ($\vec{p}_{\ell}$)
in the $J/\psi p$ ($\ell^-\bar \nu$) rest frame and 
the line of flight of the $J/\psi p$ ($\ell^-\bar \nu$) system in the $\Lambda_b$ rest frame,
while the angle $\phi$ is between the $J/\psi p$ and $\ell^-\bar \nu$ planes,
defined by the momenta of the $J/\psi p$ and $\ell^-\bar \nu$ pairs, respectively,
in the $\Lambda_b$ rest frame.
The partial decay width reads~\cite{semi_baryonic}
\begin{eqnarray}\label{4body}
d\Gamma=\frac{|\bar {\cal A}|^2}{4(4\pi)^6 m_{\Lambda_b}^3}X\beta_H\beta_L\, 
ds\, dt\, d\text{cos}\,\theta_H\, d\text{cos}\,\theta_L\, d\phi\,,
\end{eqnarray}
where $X$, $\beta_H$, and $\beta_L$ are given by
\begin{eqnarray}
X&=&\bigg[\frac{1}{4}(m_{\Lambda_b}^2-s-t)^2-st\bigg]^{1/2}\,,\nonumber\\
\beta_H&=&\frac{1}{t}\lambda^{1/2}(t,m_{J/\psi}^2,m_{p}^2)\,,\nonumber\\
\beta_L&=&\frac{1}{s}\lambda^{1/2}(s,m_{\ell}^2,m_{\bar \nu}^2)\,,
\end{eqnarray}
respectively,
with $\lambda(a,b,c)=a^2+b^2+c^2-2ab-2bc-2ca$.
The regions for the five variables of the phase space are given by
\begin{eqnarray}
(m_\ell+m_{\bar \nu})^2&\leq s\leq& (m_{\Lambda_b}-\sqrt{t})^2\,,\;\;\nonumber\\
(m_{J/\psi}+m_{p})^2    &\leq t\leq& (m_{\Lambda_b}-m_\ell-m_{\bar \nu})^2\,,\nonumber\\
0\leq \theta_{H,L}&\leq \pi\,,\,0\leq& \phi\leq 2\pi\,.
\end{eqnarray}

\section{Numerical Results and Discussions}
\begin{figure}[t!]
\centering
\includegraphics[width=2.5in]{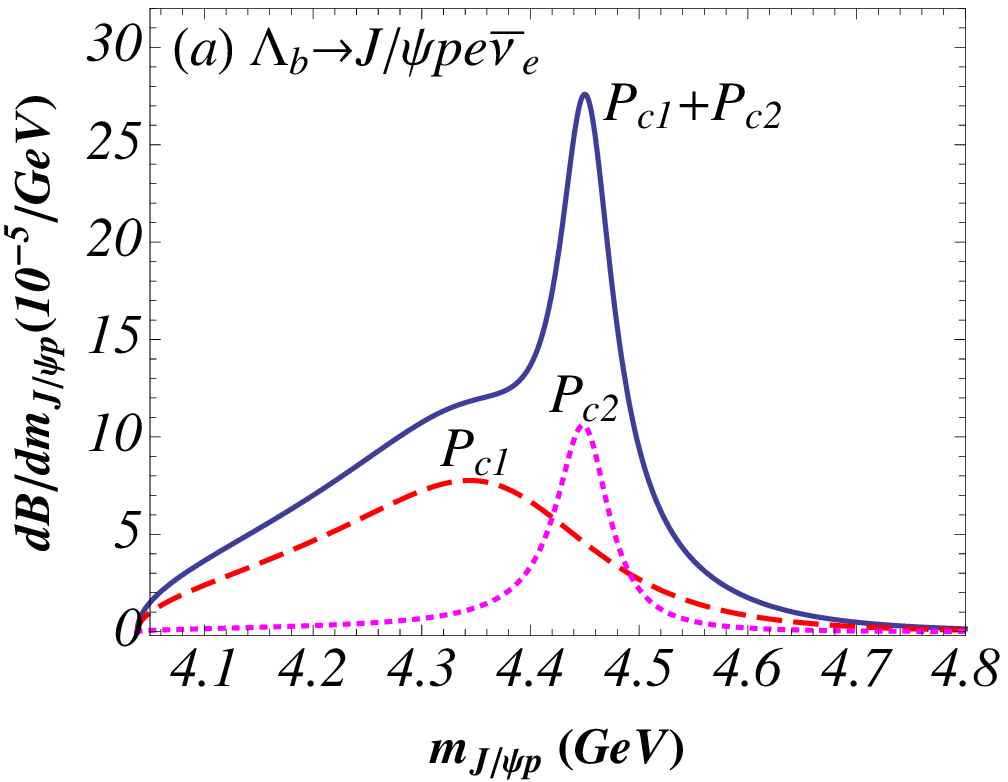}
\includegraphics[width=2.5in]{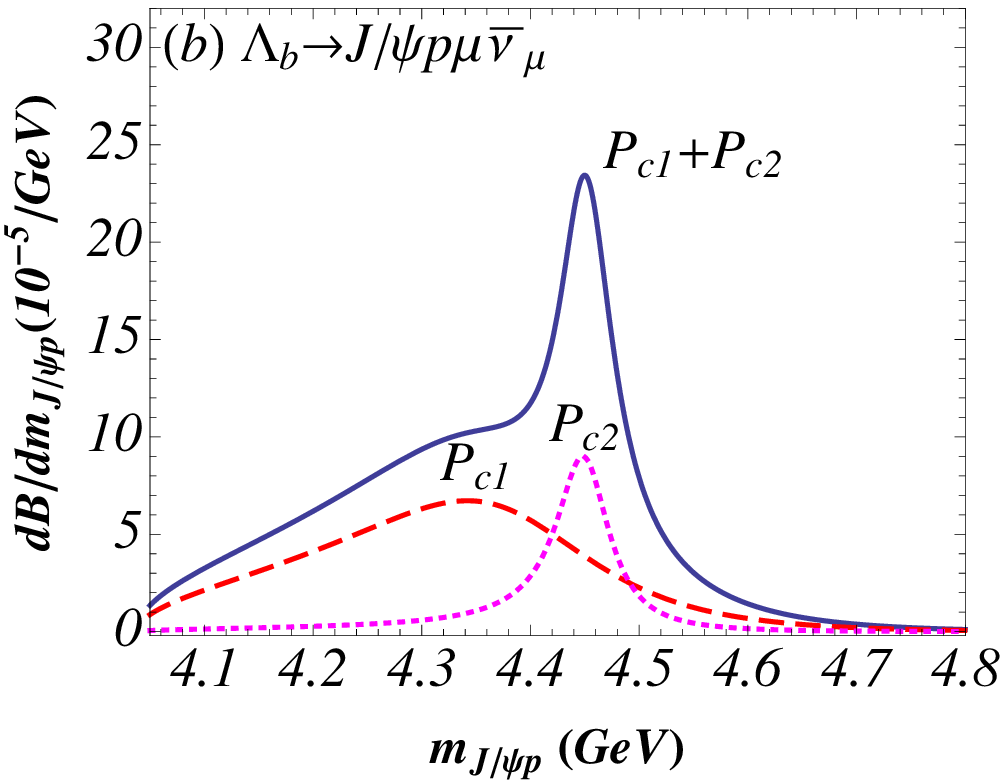}
\includegraphics[width=2.5in]{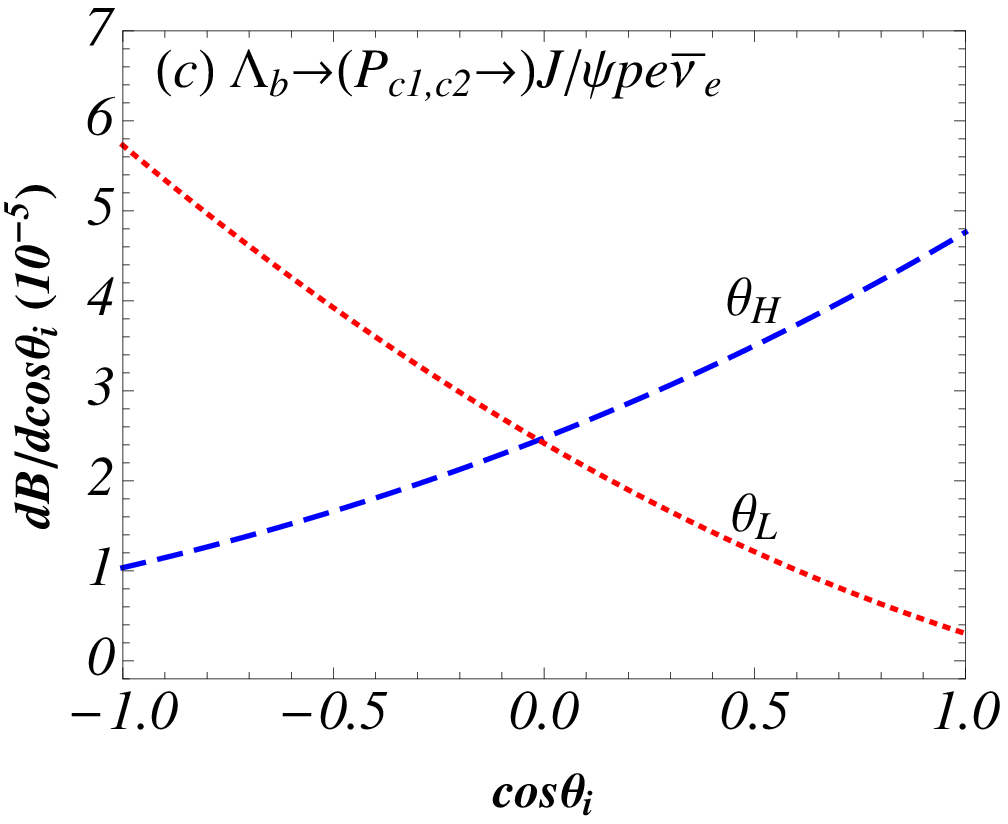}
\includegraphics[width=2.5in]{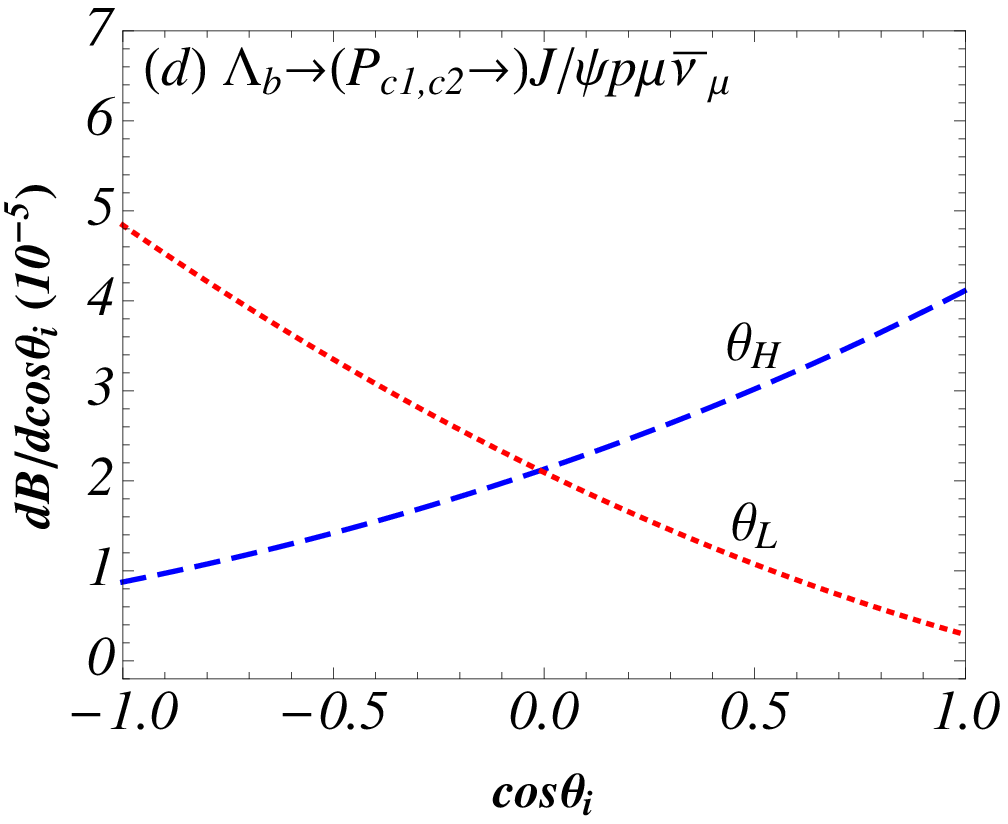}
\caption{In the first line, the $m_{J/\psi p}$ spectra of 
(a) $\Lambda_b\to J/\psi p e\bar \nu_e$ and 
(b) $\Lambda_b\to J/\psi p\mu\bar \nu_\mu$ are drawn with
the dash, dotted, and solid lines for the
${\cal P}_{c1}\to J/\psi p$, ${\cal P}_{c2}\to J/\psi p$, and 
interfering ${\cal P}_{c1,c2}\to J/\psi p$ decays.
In the second line, the angular distribution of
(c) $\Lambda_b\to ({\cal P}_{c1,c2}\to)J/\psi p e\bar \nu_e$ and 
(d) $\Lambda_b\to ({\cal P}_{c1,c2}\to)J/\psi p\mu\bar \nu_\mu$
are drawn with
the dash and dotted lines for cos$\theta_H$ and cos$\theta_L$, respectively.
}\label{dia3}
\end{figure}

For the numerical analysis, in the Wolfenstein parameterization 
$V_{ub}=A\lambda^3(\rho-i\eta)$ is presented with
$(\lambda,\,A,\,\rho,\,\eta)=(0.225,\,0.814,\,0.120\pm 0.022,\,0.362\pm 0.013)$~\cite{pdg}.
The masses and the decay widths of ${\cal P}_{c1,c2}$ are measured as~\cite{Pc_LHCb}
\begin{eqnarray}
(m_{{\cal P}_{c1}},\Gamma_{{\cal P}_{c1}})&=&(4380\pm 8\pm 29,\,205\pm 18\pm 86)\;\text{MeV},\nonumber\\
(m_{{\cal P}_{c2}},\Gamma_{{\cal P}_{c2}})&=&(4449.8\pm 1.7\pm 2.5,\,39\pm 5\pm 19)\;\text{MeV}.
\end{eqnarray}
 In equation of motion, $\bar F_{{\cal P}_{c1,c2}}$ ($\bar G_{{\cal P}_{c1,c2}}$)
via the (axial)vector currents
can be related to $F_{{\cal P}_{c1,c2}}$ defined in Ref.~\cite{Hsiao:2015nna} 
as the form factors of the resonant $\Lambda_b\to {\cal P}_{c1,c2}\to J/\psi p$ transitions
in Figs.~\ref{dia}c and \ref{dia}d via the (pseudo)scalar currents, given by
\begin{eqnarray}
\bar F_{{\cal P}_{c1,c2}}=\frac{m_b-m_s}{m_{\Lambda_b}-m_p}F_{{\cal P}_{c1,c2}}\,,\;
\bar G_{{\cal P}_{c1,c2}}=\frac{m_b+m_s}{m_{\Lambda_b}+m_p}F_{{\cal P}_{c1,c2}}\,.
\end{eqnarray}
The new observations of 
${\cal R}_{\pi K}({\cal P}_{c1},{\cal P}_{c2})=(0.050\pm 0.039,0.033\pm 0.021)$
in Eq.~(\ref{data1}) favor the nonfactorizable diagram in Fig.~\ref{dia}b 
as the main contribution. Nonetheless, due to the large uncertainties,
it still leaves the room for the charmless $b\to u\bar u q$ transitions in Figs.~\ref{dia}c and \ref{dia}d,
such that the new extractions of $F_{{\cal P}_{c1,c2}}$ for
the resonant $\Lambda_b\to {\cal P}_{c1,c2}\to J/\psi p$ transitions are needed.
Subsequently, we follow the numerical analysis in Ref.~\cite{Hsiao:2015nna}
to perform the fitting, where
the previously ignored contribution from diagram in Fig.~\ref{dia}c has been taken into account,
such that 
we obtain $F_{{\cal P}_{c1}}=2.8\pm 4.3$ and $F_{{\cal P}_{c2}}=1.6\pm 1.5$.
Note that the form factors $F_{P_{c1,c2}}$ correspond to the amount of the intrinsic charms.
According to Ref.~\cite{Brodsky:2001yt},
the intrinsic charm (IC) components are estimated to be
$1\%$, $4\%$ and $ >4\%$ within the proton, $B$ meson 
and $\Lambda_b$ baryon, respectively,
such that
the $\Lambda_b\to p J/\psi (K^-,\pi^-)$ and $p J/\psi\ell\bar \nu_\ell$ decays
with both $\Lambda_b$ and $p$
are more favourable to produce the pentaquark states. 
However, due to the theoretical difficulty, 
caused by the $c\bar c$ pair added to produce the ${\cal P}_{c1,c2}$ states,
the current calculations based on QCD models cannot 
directly relate $F_{{\cal P}_{c1,c2}}$ to the IC components yet.
Fortunately, the currently improved data allow us to extract $F_{P_{c1}}$ and $F_{P_{c2}}$,
though the newly extracted values are smaller than the ones in Ref.~\cite{Hsiao:2015nna}.
We hence predict the branching ratios and the $m_{J/\psi p}$ spectra for 
 $\Lambda_b\to J/\psi p \ell^-\bar \nu_\ell$  in  Table~\ref{tab1} and Fig.~\ref{dia3}, respectively.
Note that the tau lepton mode is forbidden 
due to $m_{\Lambda_b}<m_{J/\psi p}+m_{\tau^-\bar \nu_\tau}$. From Eq.~(\ref{4body}), 
we can evaluate the integrated angular distribution asymmetries, defined by
\begin{eqnarray}
{\cal A}_{\theta_i}\equiv
\frac{\int^1_0 \frac{d{\cal B}}{dcos\theta_i}dcos\theta_i-\int^0_{-1}\frac{ d{\cal B}}{dcos\theta_i}dcos\theta_i}
{\int^1_0 \frac{d{\cal B}}{dcos\theta_i}dcos
\theta_i+\int^0_{-1}\frac{ d{\cal B}}{dcos\theta_i}dcos\theta_i}\,,\ (i=H,\, L)\,.\label{AS}
\end{eqnarray}
Our numerical results are given in Table~\ref{tab1}, 
which are in accordance with the angular distributions in Figs.~\ref{dia3}c and \ref{dia3}d.
\begin{table}[b]
\caption{The numerical results of the branching ratios and 
the angular distribution asymmetries for 
the semileptonic $\Lambda_b\to J/\psi p\ell^-\bar \nu_\ell$ decays, 
where 
the first and second errors for ${\cal B}$ are from
the form factors and $|V_{ub}|$, respectively,
while those for
${\cal A}_{\theta_{H,L}}$  are from the form factors.
}\label{tab1}
\begin{tabular}{|c||ccc|}
\hline
modes  &  ${\cal B}\times 10^{6}$ & ${\cal A}_{\theta_H}$&${\cal A}_{\theta_L}$\\\hline
$\Lambda_b\to ({\cal P}_{c1}\to)J/\psi p e^-\bar \nu_e$
&$0.49^{+2.71}_{-0.49}\pm 0.04$&33.2\%&-51.9\%\\
$\Lambda_b\to ({\cal P}_{c2}\to)J/\psi p e^-\bar \nu_e$
&$0.81^{+2.20}_{-0.80}\pm 0.06$&38.7\%&-51.6\%\\
$\Lambda_b\to ({\cal P}_{c1,c2}\to)J/\psi p e^-\bar \nu_e$
&$2.04^{+4.82}_{-1.56}\pm 0.15$&$(36.5^{+2.6}_{-3.2})\%$&$(-51.7^{+0.2}_{-0.2})\%$\\
\hline
$\Lambda_b\to ({\cal P}_{c1}\to)J/\psi p \mu^-\bar \nu_\mu$
&$0.45^{+2.34}_{-0.45}\pm 0.03$&33.5\%&-50.7\%\\
$\Lambda_b\to ({\cal P}_{c2}\to)J/\psi p  \mu^-\bar \nu_\mu$
&$0.69^{+1.88}_{-0.68}\pm 0.05$&39.2\%&-50.2\%\\
$\Lambda_b\to ({\cal P}_{c1,c2}\to)J/\psi p  \mu^-\bar \nu_\mu$
&$1.75^{+4.14}_{-1.34}\pm 0.13$&$(37.0^{+2.7}_{-3.3})\%$&$(-50.4^{+0.2}_{-0.3})\%$\\\hline
\end{tabular}
\end{table}

In Table~\ref{tab1}, the $\Lambda_b\to J/\psi p\ell^-\bar \nu_\ell$ decays
with the single ${\cal P}_{c1}$ and ${\cal P}_{c2}$ and 
the combined ${\cal P}_{c1,c2}$ contributions all present the branching ratios 
of order $10^{-7}-10^{-6}$, which are accessible to the measurements at the LHCb.
Because of the constructive interference between the two pentaquark states,
${\cal B}(\Lambda_b\to ({\cal P}_{c1,c2}\to)J/\psi p\ell^-\bar \nu_\ell)>$
${\cal B}(\Lambda_b\to ({\cal P}_{c1}\to)J/\psi p\ell^-\bar \nu_\ell)$+
${\cal B}(\Lambda_b\to ({\cal P}_{c2}\to)J/\psi p\ell^-\bar \nu_\ell)$,
as illustrated in Figs.~\ref{dia3}a and \ref{dia3}b.
Note that  
${\cal B}(\Lambda_b\to J/\psi p e^-\bar \nu_e)
>{\cal B}(\Lambda_b\to J/\psi p \mu^-\bar \nu_\mu)$ is due to the fact 
that
$m_e<m_\mu$ allows a more  phase space for the integration.
We note that, although the $c\bar c$ formation leads to 
the suppression of ${\cal O}(1/m_c^4)$ for the probability~\cite{Brodsky:2001yt},
which is estimated in QCD models, 
the nonperturbative effects may exist to compensate the short contribution.
For example, the intrinsic charms are due to the quantum fluctuation within the hadron,
which cause a less suppression of $1/m_c^2$~\cite{Brodsky:1991dj}.
Therefore,
since the so-called hidden charm ${\cal P}_{c1,c2}$ pentaquark states 
have the additional $c\bar c$ quark pair compared to the proton,
it is reasonable to see that 
${\cal B}(\Lambda_b\to p \mu^-\bar \nu_\mu)=(4.1\pm 1.0)\times 10^{-4}$~\cite{LHCb_Vub} 
is $O(10^2)$ larger than those of ${\cal B}(\Lambda_b\to ({\cal P}_{c1,c2}\to)J/\psi p \ell^-\bar \nu_\ell)$.
The angular distribution asymmetries in Table~\ref{tab1}
demonstrates that the $J/\psi p$ and $\ell^-\bar \nu_\ell$ pairs are highly polarized.
We remark that the asymmetry of ${\cal A}_{\theta_{L}}\sim -50\%$ for the $\ell^-\bar \nu_\ell$ pair is reasonable, 
since the left-handed structure of the current $\bar \ell\gamma^\mu(1-\gamma_5)\nu_\ell$ 
inherits the left-handed feature of the  $W$-boson, which is also highly polarized. 
On the other hand, with the structure of 
$\bar u_p(F_{{\cal P}_c}\gamma_\mu-G_{{\cal P}_c}\gamma_\mu\gamma_5)u_{\Lambda_b}$
in Eq.~(\ref{amp2}) for the resonant $\Lambda_b\to {\cal P}_c\to J/\psi p$ transition,
the hadronic current is left-handed, which corresponds to our model.
Note that there is no error for ${\cal A}_{\theta_{H,L}}$ with the single ${\cal P}_{c1}$ and ${\cal P}_{c2}$ contributions
since all the uncertainties have been canceled in
Eq.~(\ref{AS}). 

\section{Conclusions}
We have studied 
the semileptonic $\Lambda_b\to ({\cal P}_{c1,c2}\to)J/\psi p \ell^-\bar \nu_\ell$ decays
with the $\Lambda_b\to {\cal P}_{c1,c2}\to J/\psi p$ transition form factors as
those proposed to explain the 
pentaquark states in
the $\Lambda_b\to K^-({\cal P}_{c1,c2}\to)J/\psi p$ decays.
We have found 
that 
${\cal B}(\Lambda_b\to J/\psi p \ell^-\bar \nu_\ell)=
(2.04^{+4.82}_{-1.57},1.75^{+4.14}_{-1.35})\times 10^{-6}$ for $\ell=(e,\mu)$, 
for $\ell=(e,\mu)$, which are about 100 times smaller than 
the observed decay of  $\Lambda_b\to p\mu\bar \nu_\mu$.
We have also shown  the angular distribution asymmetries for the $J/\psi p$ and $\ell^-\bar \nu_\ell$ pairs.
Our results for the decay branching ratio and asymmetries in $\Lambda_b\to ({\cal P}_{c1,c2}\to)J/\psi p \ell^-\bar \nu_\ell$
are accessible to the experiments at the LHCb, which will clearly  improve the knowledge 
of the hidden charm pentaquark states.

\section*{ACKNOWLEDGMENTS}
The work was supported in part by 
National Center for Theoretical Sciences, 
National Science Council (NSC-101-2112-M-007-006-MY3), 
MoST (MoST-104-2112-M-007-003-MY3), and NSFC (11675030 and 11547008).



\begin{thebibliography}{99}
\bibitem{Pc_LHCb}
R.~Aaij {\it et al.} [LHCb Collaboration], Phys.\ Rev.\ Lett.\  {\bf 115}, 072001 (2015).  

\bibitem{Pc_LHCb2}
R.~Aaij {\it et al.} [LHCb Collaboration], Phys.\ Rev.\ Lett.\  {\bf 117}, 082002 (2016).

\bibitem{Aaij:2016ymb} 
R.~Aaij {\it et al.} [LHCb Collaboration],
Phys.\ Rev.\ Lett.\  {\bf 117}, 082003 (2016).

\bibitem{Wu:2010jy} 
J.J.~Wu, R.~Molina, E.~Oset and B.S.~Zou,
Phys.\ Rev.\ Lett.\  {\bf 105}, 232001 (2010).

\bibitem{Karliner:2015ina} 
M.~Karliner and J.L.~Rosner,
Phys.\ Rev.\ Lett.\  {\bf 115}, 122001 (2015).

\bibitem{Yuan:2012wz} 
S.G.~Yuan, K.W.~Wei, J.~He, H.S.~Xu and B.S.~Zou,
Eur.\ Phys.\ J.\ A {\bf 48}, 61 (2012).

\bibitem{Chen:2015loa} 
R.~Chen, X.~Liu, X.Q.~Li and S.L.~Zhu,
Phys.\ Rev.\ Lett.\  {\bf 115}, 132002 (2015).

\bibitem{Chen:2016qju} 
For a review on the pentaquark states, see H.X.~Chen, W.~Chen, X.~Liu and S.L.~Zhu,
Phys.\ Rept.\  {\bf 639}, 1 (2016).
\bibitem{Wang:2015epa} 
Z.G.~Wang, Eur.\ Phys.\ J.\ C {\bf 76}, 70 (2016).

\bibitem{Maiani:2015vwa} 
L.~Maiani, A.D.~Polosa and V.~Riquer, Phys.\ Lett.\ B {\bf 749}, 289 (2015).

\bibitem{Anisovich:2015cia} 
V.V.~Anisovich, M.A.~Matveev, J.~Nyiri, A.V.~Sarantsev and A.N.~Semenova,
arXiv:1507.07652 [hep-ph].

\bibitem{Ghosh:2015ksa} 
R.~Ghosh, A.~Bhattacharya and B.~Chakrabarti, arXiv:1508.00356 [hep-ph].


\bibitem{Chen:2015moa} 
H.X.~Chen, W.~Chen, X.~Liu, T.G.~Steele and S.L.~Zhu, 
Phys.\ Rev.\ Lett.\  {\bf 115}, 172001 (2015).

\bibitem{Roca:2015dva} 
L.~Roca, J.~Nieves and E.~Oset, Phys.\ Rev.\ D {\bf 92}, 094003 (2015).

\bibitem{Meissner:2015mza} 
U.G.~Mei§ner and J.A.~Oller, Phys.\ Lett.\ B {\bf 751}, 59 (2015).

\bibitem{Burns:2015dwa} 
T.J.~Burns, Eur.\ Phys.\ J.\ A {\bf 51}, 152 (2015).

\bibitem{Lu:2016nnt} 
Q.F.~L and Y.B.~Dong, Phys.\ Rev.\ D {\bf 93}, 074020 (2016).


\bibitem{Mironov:2015ica} 
A.~Mironov and A.~Morozov, 
JETP Lett.\  {\bf 102}, 271 (2015).

\bibitem{Guo:2015umn} 
F.~K.~Guo, U.~G.~Mei$\beta$ner, W.~Wang and Z.~Yang,
Phys.\ Rev.\ D {\bf 92}, 071502 (2015).

\bibitem{Liu:2015fea} 
X.~H.~Liu, Q.~Wang and Q.~Zhao,
Phys.\ Lett.\ B {\bf 757}, 231 (2016).


\bibitem{Mikhasenko:2015vca} 
M.~Mikhasenko,
arXiv:1507.06552 [hep-ph].


\bibitem{Wang:2015pcn} 
E.~Wang, H.X.~Chen, L.S.~Geng, D.M.~Li and E.~Oset,
Phys.\ Rev.\ D {\bf 93}, 094001 (2016).
\bibitem{Li:2015gta} 
G.N.~Li, X.G.~He and M.~He, JHEP {\bf 1512}, 128 (2015);
H.Y.~Cheng and C.K.~Chua, Phys.\ Rev.\ D {\bf 92}, 096009 (2015).

\bibitem{Hsiao:2015nna} 
Y.K.~Hsiao and C.Q.~Geng, 
Phys.\ Lett.\ B {\bf 751}, 572 (2015).

\bibitem{Wang:2015jsa} 
Q.~Wang, X.H.~Liu and Q.~Zhao,
 Phys.\ Rev.\ D {\bf 92}, 034022 (2015).

\bibitem{Kubarovsky:2015aaa} 
V.~Kubarovsky and M.B.~Voloshin,
Phys.\ Rev.\ D {\bf 92}, 031502 (2015).


\bibitem{Karliner:2015voa} 
M.~Karliner and J.L.~Rosner,
Phys.\ Lett.\ B {\bf 752}, 329 (2016).

\bibitem{Lu:2015fva} 
Q.F.~Lü, X.Y.~Wang, J.J.~Xie, X.R.~Chen and Y.B.~Dong,
Phys.\ Rev.\ D {\bf 93}, 034009 (2016)


\bibitem{Aaij:2015fea} 
R.~Aaij {\it et al.} [LHCb Collaboration], Chin.\ Phys.\ C {\bf 40}, 011001 (2016).



\bibitem{Lebed:2015dca} 
R.F.~Lebed, Phys.\ Rev.\ D {\bf 92}, 114030 (2015).



\bibitem{IC1}
S.J.~Brodsky, P.~Hoyer, C.~Peterson and N.~Sakai, Phys.\ Lett.\ B {\bf 93}, 451 (1980).

\bibitem{IC2}
S.J.~Brodsky, C.~Peterson and N.~Sakai, Phys.\ Rev.\ D {\bf 23}, 2745 (1981).

\bibitem{Brodsky:1997fj} 
S.J.~Brodsky and M.~Karliner, Phys.\ Rev.\ Lett.\  {\bf 78}, 4682 (1997).

\bibitem{Mikhasenko:2012km} 
M.~Mikhasenko, 
Phys.\ Atom.\ Nucl.\  {\bf 77}, 623 (2014) [Yad.\ Fiz.\  {\bf 77}, 658 (2014)].

\bibitem{Chang:2001iy} 
C.H.V.~Chang and W.S.~Hou, Phys.\ Rev.\ D {\bf 64}, 071501 (2001).

\bibitem{Brodsky:2001yt} 
S.J.~Brodsky and S.~Gardner, Phys.\ Rev.\ D {\bf 65}, 054016 (2002).

\bibitem{LHCb1} 
R.~Aaij {\it et al.} [LHCb Collaboration], JHEP {\bf 1407}, 103 (2014).

\bibitem{Wang:2015rcz} 
W.~Wang and Q.~Zhao, 
Phys.\ Lett.\ B {\bf 755}, 261 (2016).

\bibitem{Hsiao:2016pml} 
Y.K.~Hsiao and C.Q.~Geng,
Chin.\ Phys.\ C {\bf 41}, 013101 (2017).

\bibitem{Brodsky:2001yt} 
S.J.~Brodsky and S.~Gardner,
Phys.\ Rev.\ D {\bf 65}, 054016 (2002).

\bibitem{Brodsky:1991dj} 
S.J.~Brodsky, P.~Hoyer, A.H.~Mueller and W.K.~Tang,
Nucl.\ Phys.\ B {\bf 369}, 519 (1992).


\bibitem{LHCb_Vub}
R.~Aaij {\it et al.} [LHCb Collaboration], Nature Phys.\ {\bf 11}, 743 (2015). 

\bibitem{Chen:2002zk} 
C.H.~Chen and C.Q.~Geng,
Phys.\ Rev.\ D {\bf 66}, 094018 (2002).


\bibitem{Gabbiani:2002ti} 
F.~Gabbiani, J.W.~Qiu and G.~Valencia, 
Phys.\ Rev.\ D {\bf 66}, 114015 (2002). 

\bibitem{pdg}
K.A.~Olive {\it et al.}  [Particle Data Group Collaboration], Chin.\ Phys.\ C {\bf 38}, 090001 (2014).

\bibitem{Kl4}
A. Pais and S.B. Treiman, Phys. Rev. {\bf 168}, 1858 (1968).

\bibitem{Wise}
C.L.Y.~Lee, M.~Lu and M.B.~Wise, Phys.\ Rev.\  D {\bf 46}, 5040 (1992).

\bibitem{semi_baryonic} 
C.Q.~Geng and Y.K.~Hsiao, Phys.\ Lett.\ B {\bf 704}, 495 (2011);
Phys.\ Rev.\ D {\bf 85}, 094019 (2012).

\end{thebibliography}
\end{document}